\def\deg{^{\circ}}
\def\etal{{\it et al.}}
\def\ie{{\it i.e.}}
\def\eg{{\it e.g.}}
\def\cf{{\it cf.}}
\documentclass[useAMS]{mn2e}
\usepackage{graphicx}
\usepackage{bm}

\def\eq{\begin{equation}}
\def\en{\end{equation}}
\def\lab{\label}

\def\apj{{\it Ap.J.}\thinspace}

\def\aap{{\it A\&A}\thinspace}
\def\mnras{{\it MNRAS}\thinspace}
\def\P3hat{{\mathaccent 94 P}_3}

\title[Polar cap model for bi-drifting]{Pulsar bi-drifting: implications for polar cap geometry}

\author[Wright \& Weltevrede]
{Geoff Wright$^{1}$\thanks{E-mail: Geoffrey.Wright@manchester.ac.uk} and Patrick Weltevrede$^{1}$\\
$^{1}$Jodrell Bank Centre for Astrophysics, Alan Turing Building, University of Manchester, Manchester, M13 9PL, UK}

\begin{document}

\date{Accepted 2016 month day. Received 2016 month day; in original form 2016 month day}

\maketitle

\label{firstpage}

\begin{abstract}
For many years it has been considered puzzling how pulsar radio emission, supposedly created by a circulating carousel of sub-beams, can produce the driftbands demonstrated by PSR J0815+09, and more recently PSR B1839--04, which simultaneously drift in opposing directions. Here we suggest  that the carousels of these pulsars, and hence their beams, are not circular but elliptical with axes tilted with respect to the fiducial plane. We show that certain relatively unusual lines of sight can cause bi-drifting to be observed, and a simulation of the two known exemplars is presented. Although bi-drifting is rare, non-circular beams may be common among pulsars and reveal themselves by having profile centroids displaced from the fiducial plane identified by polarisation position angle swings. They may also result in profiles with asymmetric and frequency-dependent component evolution. It is further suggested that the carousels may change their tilt by specific amounts and later reverse them. This may occur suddenly, accompanying a mode change ({\eg} PSR B0943+10), or more gradually and short-lived as in ``flare'' pulsars ({\eg} PSR B1859+07). A range of pulsar behaviour ({\eg} the shifting drift patterns of PSRs B0818--41 and B0826--34) may also be the result of non-circular carousels with varying orientation. The underlying nature of these carousels -- whether they are exclusively generated by polar cap physics or driven by magnetospheric effects -- is briefly discussed.         
\end{abstract}

\maketitle

\section{Introduction}

The phenomenon of bi-drifting has been seen as a challenge to the carousel model for polar cap emission, first proposed by Ruderman \& Sutherland in 1975. In this popular model emission in the form of drifting subpulses arises through the rotation of  ``sparks'' located on the pulsar's polar cap. The rotation is due to ${\bm{E}\times {\bm B}}$ drift on open magnetic field lines, and represents the failure of the plasma to precisely corotate with the star, so the drift around the polar cap is opposite in sense to the rotation of the pulsar. By its nature, in the aligned geometry assumed by Ruderman \& Sutherland the drift can only have one sense around the magnetic pole (otherwise the sparks and associated emission would rotate faster than the star itself). However, the rate of drift may vary with the degree of discharge of the electric potential difference.

In an inclined pulsar the \emph{observed} drift direction in the pulse window will depend on whether the observer's line of sight crosses the carousel within or outside the angle between the pulsar's rotation and magnetic axes (Ruderman 1976). If the carousel is circular (or symmetric about the plane containing both the rotational and magnetic axes  -- the fiducial plane) only one sense will be observed in both components (or in all four components if we are dealing with two nested carousels). In most pulsars which exhibit subpulse drift the observed sense is always in the same direction in a given pulsar (Weltevrede {\etal} 2006, 2007a). Exceptionally, examples have been found ({\eg} PSR B2303+30, Redman {\etal} 2005) where the drift appears to reverse, but does this in all components at the same time and can be argued to arise from aliasing, {\ie} a consequence of the observer's inability to sample the rapidly-moving drift sufficiently often and thereby giving the illusion of reverse drift (Gupta {\etal} 2004). 

In contrast, ``bi-drifting'', whereby the subpulses in different components move \emph{simultaneously} in opposing directions, has not found an easy explanation. It is a feature originally observed uniquely in the four-component pulsar J0815+09 (McLaughlin {\etal} 2004) but later also in PSR B1839--04 (Weltevrede {\etal} 2006).  A description of the properties of PSR B1839--04 are given in Weltevrede 2016 (henceforth paper I), where the difficulties for modelling the bi-drift emission are described in detail. 

PSRs J0815+09 and B1839--04 have rotational periods which are slightly above average (0.85 s and 1.8 s respectively) and inferred magnetic field strengths somewhat below average, (0.8 and 0.1 $\times10^{12}$ G respectively), but these properties are in no way unusual among pulsars. There is therefore no suggestion that bi-drifting is the result of exceptional physical conditions. Furthermore, for many pulsars -- for example in the detailed work on PSR B0943+10 ({\eg} Deshpande \& Rankin 2001, Suleymanova \& Rankin 2009, Backus {\etal} 2011, Bilous {\etal} 2014) -- the assumption of a perfectly circular carousel has proved capable of giving insight into the likely number of sparks and their rotation speed.  

However, the imposition of precise circularity has always been a mathematical convenience, requiring a precise dipolar magnetic field surrounding the magnetic pole, whereas it is known that the Ruderman \& Sutherland model requires the presence of multipoles to function as a source of pair-creation. There is then no reason why the ${\bm{E}\times {\bm B}}$ drift would be perpendicular to the dipolar radial vector. This suggests that circularity is simply an  approximation, even if perhaps a good one, to the real polar cap condition\footnote{Gil \& Sendyk (2003) acknowledge this when modelling the apparently well-defined circulation of B0943+10's carousel.}.

If the existence of a carousel can be accepted, then as an alternative to a circular form we have sought a geometry which to first order has all the basic features of a circle: a closed convex shape centred on the magnetic pole.  An ellipse was chosen, not only because it has these features, but also from a practical point of view it can be dealt with more easily mathematically and approximates naturally to a circle when its eccentricity $e$ is close to zero. As will be seen, in most pulsars the difference may be barely detectable observationally and the ellipse, like the circle, remains an approximation reflecting some little understood underlying physics.

When an elliptical carousel is so aligned that either its major or minor axis lies in the fiducial plane, then it is immediately obvious, as with the circular case, that bi-drifting cannot be observed. However, an elliptical carousel may be \emph{tilted} with respect to the fiducial plane. As is formally shown in the next section, this allows bi-drifting to be observed at least for a limited range of sightlines. There is no \emph{a priori} physical reason why a pulsar's carousel has any particular orientation with respect to our line of sight or be centred on the magnetic pole, and indeed, as we estimate below, carousels with extreme eccentricity and tilt are unlikely to be common -- otherwise bi-drifitng would be more frequently observed. 

To model the emission of the two known bi-drifting pulsars, both of which have four outer components, we need to consider two nested beams/carousels, the presence of which has been plausibly argued for many pulsars with M and D classification by Rankin (1993). For convenience, we assume here for each pulsar that the inner and outer carousels have identical geometry ({\ie} they are ellipses with the same eccentricity and common major axis, differing only in their radii). 

However, we would stress that in reality the divergencies from circular carousels (and emission beams) need not follow our precise mathematics, and, further, the inner carousel may have a different form to the outer. The purpose of this paper is to demonstrate that by permitting near-circular, somewhat egg-shaped, carousels to exist, bi-drifting and a range of other apparently puzzling observational features may be explained without necessarily requiring any additional physics.

In the next section we present the formal calculation based on a tilted elliptical beam, which we show to be capable of displaying bi-drifting. In Section 3 we discuss how a tilted beam might explain the two known bi-drifters and in Section 4 a wider range of pulsar phenomena. In Section 5 we consider the observational effects, including mode-changing, of allowing the carousel alignment to vary in time, and in Section 6 briefly discuss possible physical scenarios which might give rise to asymmetric carousels. In Section 7 we summarise our conclusions.

\begin{figure*}
\begin{center}
\includegraphics[height=0.48\hsize,angle=-90,trim=90 90 90 90,clip]{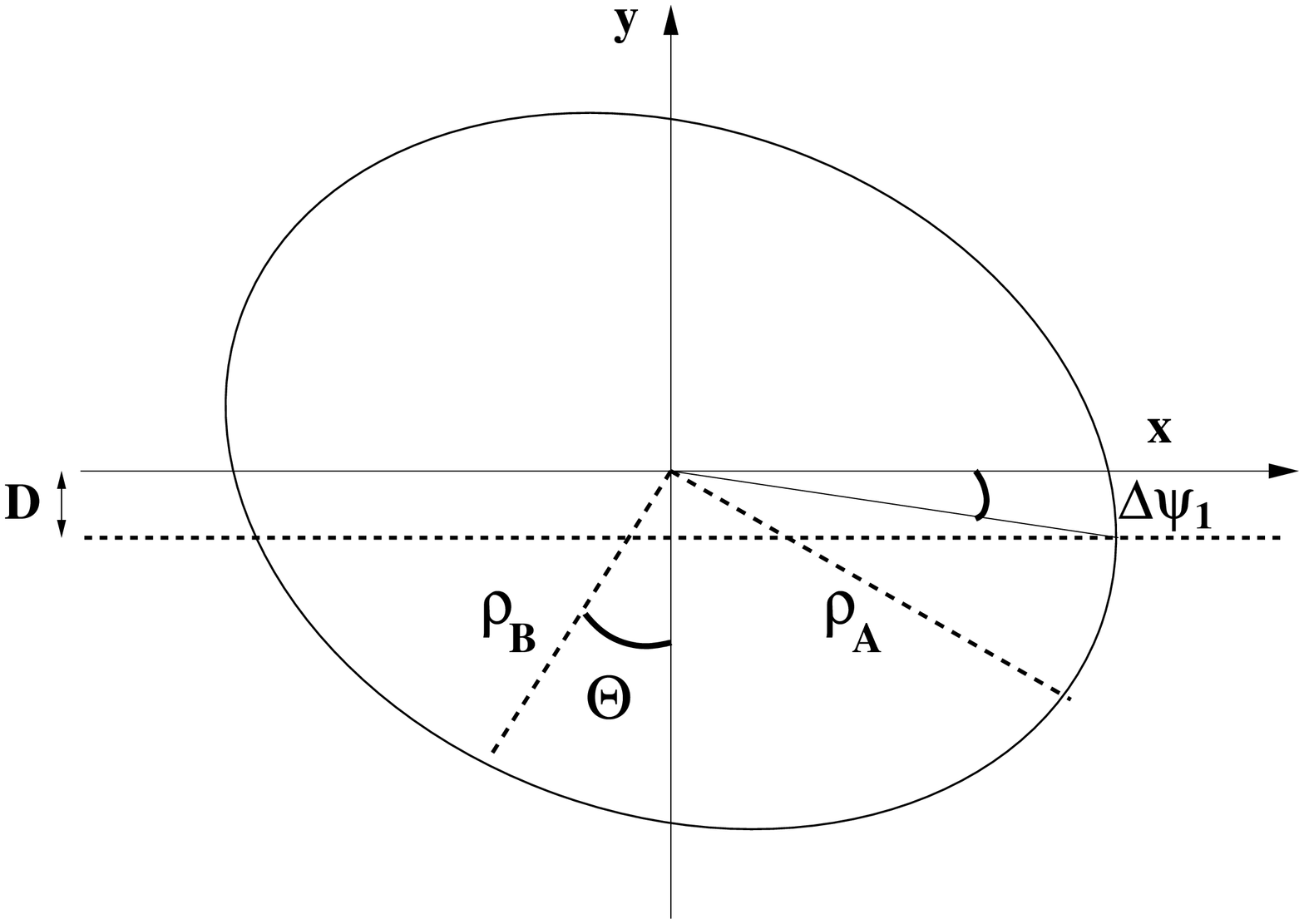}
\hspace{0.01\hsize}
\includegraphics[height=0.48\hsize,angle=-90,trim=90 90 90 90,clip]{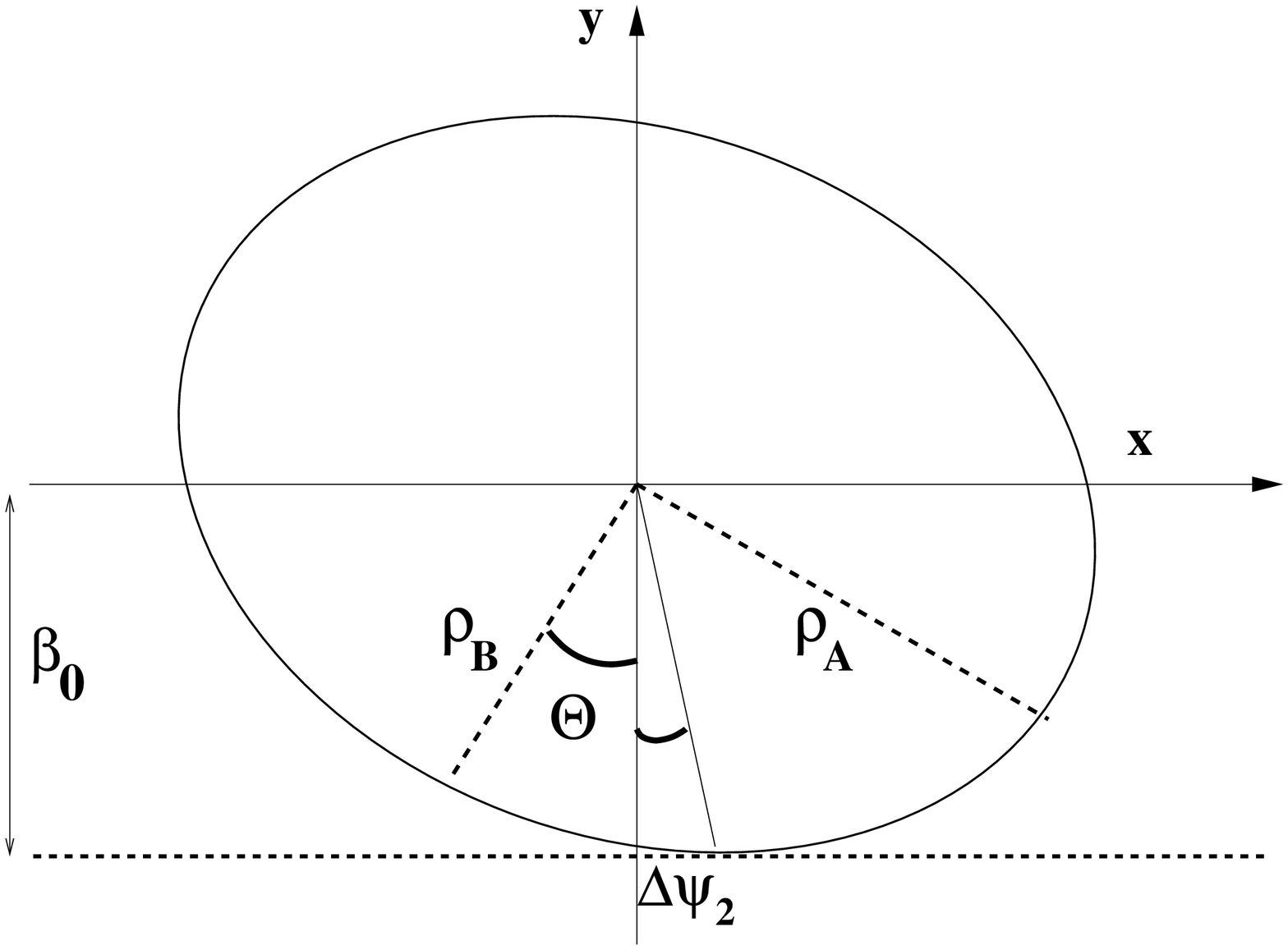}
\caption{An elliptical beam centred on the magnetic pole with major and minor axes of $\rho_A$ and $\rho_B$ respectively, and tilted at an angle $\Theta$ to the fiducial plane $x=0$. \emph{Left}: the horizontal dotted line intersects the trailing ellipse at a point where the tangent to the ellipse is vertical. If the circulation of the sparks is anti-clockwise, and the line of sight traverses horizontally from left to right within the space $0>y>-D$, then drift in the leading component will be earlier to later and in the trailing component later to earlier. Hence bi-drifting will be observed. \emph{Right}: the maximum possible impact parameter $\beta_0$ grazes the beam at position angle $\Delta\psi_2$, which can be shown to be equivalent to the angle $ \Delta\psi_1$ in the beam on the left under the transformation $\Theta\rightarrow(90\degr-\Theta)$ (see Appendix).}
\label{Fig1}
\end{center}
\end{figure*}

\section{Geometry of a tilted ellipse}
The first  purpose of this paper is to formally demonstrate how bi-drifting can arise when an appropriate asymmetric beam is observed. The beam is assumed to be generated by a carousel of similar topography on the pulsar's polar cap. For mathematical convenience we consider the beam to have the form of an ellipse with the magnetic axis at its centre. The ellipse's minor axis is tilted by an angle $\Theta$ from the fiducial plane\footnote{If the major axis were tilted $\Theta$ to the fiducial plane, it would be equivalent to transforming $\Theta\rightarrow90\degr-\Theta$ in the formulae given here.} (Fig.~\ref{Fig1}). Before tilting, the equation of the ellipse expressed in terms of ($x,y$), the beam's azimuthal and co-latitudinal coordinates respectively, is
\eq
\lab{Eq1}
\frac{x^2}{\rho_{A}^2}+\frac{y^2}{\rho_{B}^2}=1.
\en
Here $\rho_{A}$ and $\rho_{B}$ are the dimensions of the major and minor axes respectively, so that the eccentricity $e$ is given by $\rho_{B}^2=\rho_{A}^2(1-e^2)$. Following the tilt, the point at which the tangent to the ellipse is vertical (i.e. parallel to the $y$-axis) swings by an angle $\Delta\psi_1$ to the $x$-axis to a point where $y=-D$ such that
\eq
\lab{Eq2}
D=\rho_A\frac{e^2\sin\Theta\cos\Theta}{(1-e^2\sin^2\Theta)^\frac{1}{2}}
\en
(see Fig.~\ref{Fig1}, left and Appendix). 

Let us assume a carousel of beam elements circulates anti-clockwise around the centre of the ellipse, which is consistent (Ruderman 1976) with a line of sight traversing from left to right (Fig.~\ref{Fig1}) at an impact angle $\beta$. Then any horizontal line of sight (a valid approximation for pulsars well away from alignment) moving across the lower plane below the line $y=-D$ will record subpulse drift from earlier to later pulse phase in both the leading and trailing components throughout the plane ({\ie} positive drift ``+''). However between $y=-D$ and $y=0$ the drift will be earlier to later in the leading component and later to earlier in the trailing component (negative drift ``--''). Thus ``bi-drifting'' may be observed. Clearly, the reverse phenomenon can occur in the upper half-plane ($y>0$) if $0<y<D$. Thus four different patterns may be observed: -- --, -- +, + --, ++.

The tilt means that the maximum possible impact parameter which can still observe the beam is greater than $\rho_B$ (Fig.~\ref{Fig1}, right) and is given by
\eq
\lab{Eq3}
\beta_0=\rho_A(1-e^2\cos^2\Theta)^\frac{1}{2}
\en
(see Appendix). So, for given $\Theta$ and $e$, the chance of observing bi-drifting is
\eq
\lab{Eq4}
\frac{D}{\beta_0}=\frac{e^2\sin\Theta\cos\Theta}{(1-e^2\sin^2\Theta)^\frac{1}{2}(1-e^2\cos^2\Theta)^{\frac{1}{2}}}.
\en
For fairly extreme values of asymmetry in the beam, say $\Theta=45\deg$ and $e=0.8$, this ratio is about 0.5. Thus the observed rarity of bi-drifting indicates that these values are uncommon in practice. For $\Theta=20\deg$ and $e=0.2$, for example,  the chance of detecting bi-drifting drops dramatically to 1.3\%. Thus it is quite possible for tilted beams to be widespread without bi-drifting being observed.

\begin{figure}
\begin{center}
\includegraphics[height=0.99\hsize,angle=-90,trim=90 40 90 40,clip]{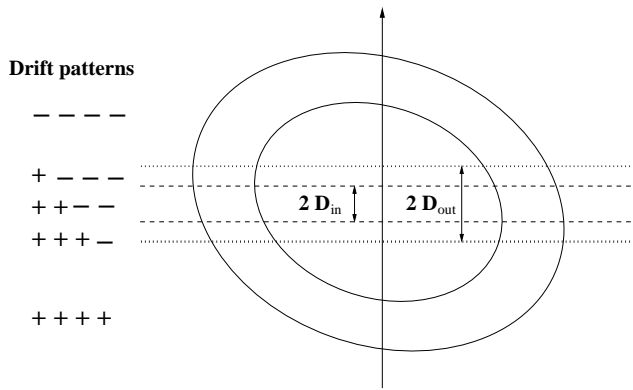}
\caption{A double cone beam consisting of two nested ellipses of identical eccentricity and tilt. In this configuration the drift direction may be observed in any of five possible combinations, shown on the left, assuming a sub-beam carousel circulating anti-clockwise, a line of sight traversing from left to right, and taking positive (+) drift from leading to trailing edges of the pulse window.}
\label{Fig2}
\end{center}
\end{figure}

Expanding this idea to account for the bi-drifting in PSR J0815+09, which has four profile components, we now consider a double-cone structure. For simplicity we assume the beam to consist of two nested ellipses of the same eccentricity, tilted by the same angle. Equation \ref{Eq2} now gives two different azimuthal bands of widths $2D_\mathrm{in}$ and $2D_\mathrm{out}$ corresponding to the major axes of the inner and outer ellipses, both of which are likely to be very small relative to the beam width $\rho_{A}$ unless the tilt angle is large and the ellipse far from circular. Then theoretically we may observe any of five different drift-patterns, shown in Fig.~\ref{Fig2}, of which three exhibit bi-drifting. A further three possible bi-drift sequences within $2D_\mathrm{out}$, reversed in drift direction to those given in Fig.~\ref{Fig2}, would also be possible if the tilt angle of the ellipse were reversed ({\ie} if the major axis ran from bottom left to upper right in the figure).

An interesting consequence of the asymmetric carousel model is that bi-drifting observed at low frequencies may disappear, or at least diminish, at higher frequencies. If a radius-to-frequency mapping is present (Cordes 1978), then for a given $\beta$ the beam is smaller as frequency increases, hence the observer's line of sight samples emission at larger $\beta/D$. Single-pulse observations of both known bi-drifting pulsars at L-band and above would be of value in testing and resolving this. 
 
As argued above, in most pulsars it is more likely that $D_\mathrm{in,out}/|\beta|$ will be small, so that the observed sense of drift will be the same across the profile -- although the rate of drift (and hence $P_2$, the drift band separation in pulse longitude) will vary from component to component and the width of the integrated components will reflect this. For example, a line of sight which intersects the carousel of Fig.~\ref{Fig1} in the upper sector where the drift is in the same sense in both components, will show the subpulses drifting across a narrower window in the leading component. The observed drift band repetition time ($P_3$) will be identical in both components, and as a consequence the drift band will be steeper in the leading component. In double-cone carousels such as Fig.~\ref{Fig2} both the leading components will be narrow and the trailing components broad and possibly merged (or the reverse if the tilt is in the opposite sense). The profiles and subpulse behaviour of PSR J1819+1305 (Rankin \& Wright 2008), and arguably PSR B1918+19 (Rankin {\etal} 2013), may result from this geometry. 

\section{Application to known bi-drifters}
In Figs.~\ref{Fig1} and~\ref{Fig2} we have assumed that the line of sight is a straight line. This is valid for angles of inclination ($\alpha$) considerably larger than $|\beta|$. For polar caps closer to the pulsar's rotation axis the line of sight's curvature must be taken into account ({\eg} Dyks \& Pierbattista 2015). The bi-drifting pulsar J0815+09 has four distinct components which occupy approximately one third of the pulsar's duty cycle. This is a strong indication that $\alpha$ is relatively small. The other known bi-drifter \mbox{PSR B1839--04} also has a wide profile (about 25\% of its duty cycle) and is shown in Paper I on polarisation grounds to be most probably of relatively low $\alpha$ with a line of sight passing between the rotation and magnetic axes. 

To reflect this more complex geometry we created a numerical simulation and experimented with
the parameters of the elliptical beam, its inclination with respect to the fiducial plane, and the sightline parameters ($e, \rho_{A},\Theta, \alpha, \beta$).  In each pulsar we attempted to mimic the observed driftband separation ($P_2$), their repetition rates ($P_3$) and the phase differences between the four components.

The simulation adopted a range of self-imposed constraints. It was assumed that the inner and outer beams took the form of mathematical ellipses with identical eccentricity $e$ and tilt $\Theta$. Each beam consisted of  the same number ($N$) of emission regions which moved around the ellipse in the same circulation time of $P_4=N \times P_3$ and with unvarying size and intensity, represented as a 2-dimensional Gaussian distribution. An arbitrary but fixed phase difference could be applied between the inner and outer beams. These constraints were driven by the need to maintain clarity in presentation and not intended to reflect any particular polar cap or emission physics. Our purpose is to show schematically that the model can work in principle. The production of a realistic fit would require more degrees of freedom, by including, for example, randomly fluctuating (non-drifting) background  emission and a specific intensity distribution for the beam. 

In respect of the circulation, two options were allowed: either the subbeams circulated the magnetic axis at a fixed angular rotation rate, which led to bunching when they were close to the magnetic pole, or they had to maintain a fixed spatial distance from one another throughout the circulation, which led to varying angular rotation rates at different points of their circulation.

\begin{figure*}
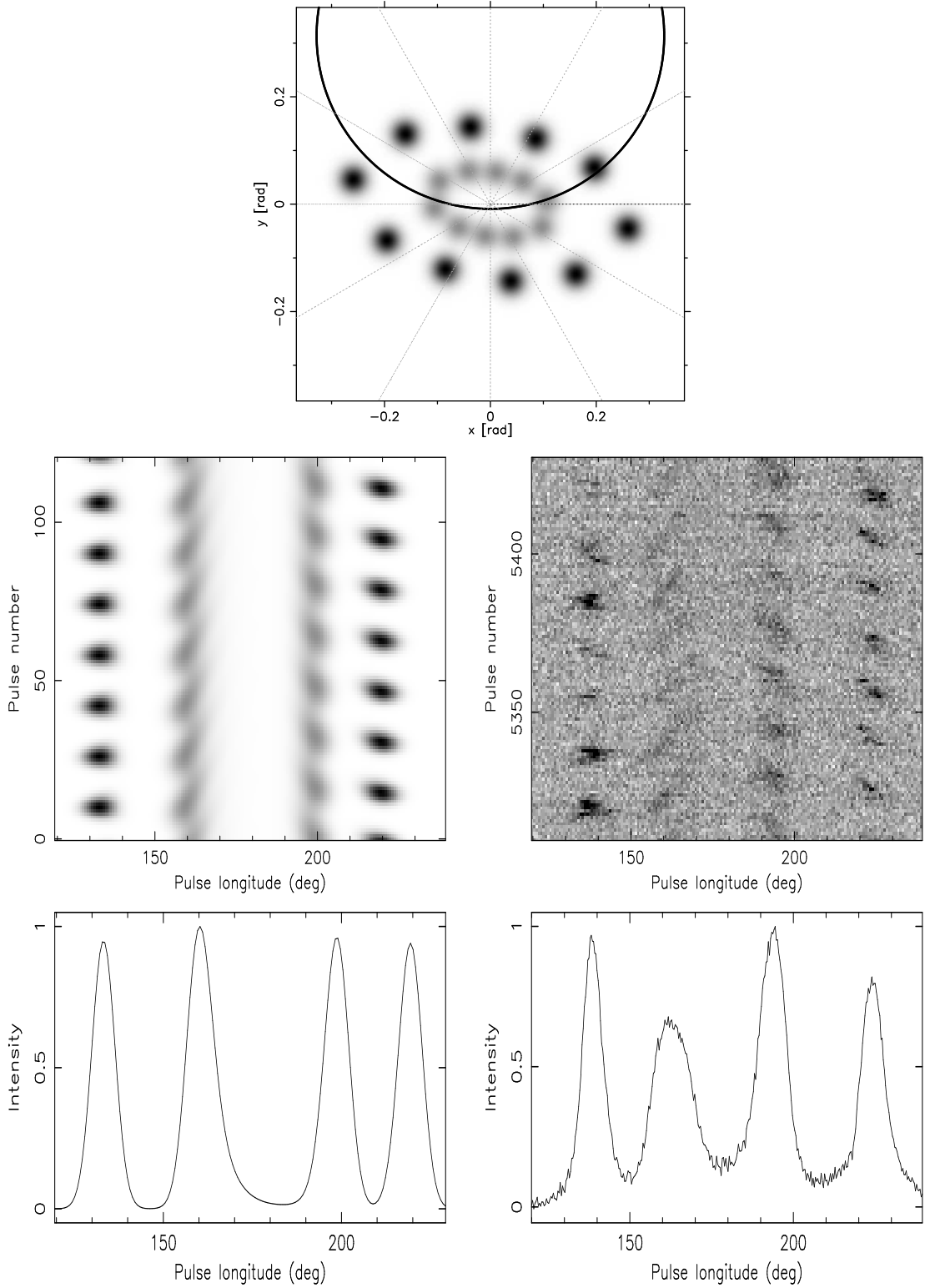

\begin{center}
\begin{tabular}{@{}l@{}l}
{\mbox{\includegraphics[width=7cm,height=7cm,angle=-90]{0815fakemapx.ps}}}\\
\end{tabular}
\begin{tabular}{@{}l@{}l}
{\mbox{\includegraphics[width=7cm,height=7cm,angle=-90]{0815fakestackx.ps}}}\ \ \ \ \ \ \ &
{\mbox{\includegraphics[width=7cm,height=7cm,angle=-90]{0815realstack.ps}}}\\
{\mbox{\includegraphics[width=6cm,height=7cm,angle=-90]{0815fakeprofilex.ps}}}\ \ \ \ \ \ \ &
{\mbox{\includegraphics[width=6cm,height=7cm,angle=-90]{0815realprofile.ps}}}\\
\end{tabular}
\caption{A comparison of the simulation of the emission of PSR J0815+09 with an archival Arecibo observation at 327 MHz. \emph{top}: two nested elliptical beams with a common eccentricity, each containing 10 sub-beams tilted at an angle of $12\deg$ to the fiducial plane rotating anticlockwise about the magnetic pole, which is inclined at $\alpha=18\deg$ to the pulsar's rotation axis and viewed with an impact parameter $\beta$ of $0.5\deg$ (resulting in the line of sight indicated by the solid line). \emph{centre}: Emission drifts in opposing senses on either side of the fiducial plane (simulation left, observed pulses right). \emph{bottom}: The resulting integrated profile (simulation left, observation right).}
\label{fig3}
\end{center}
\end{figure*}

\subsection{PSR J0815+09}
The published single pulse sequence of this pulsar's emission (McLaughlin {\etal} 2004,  Fig.~\ref{Fig1}) show the basic drift pattern of the four emission bands to be (0, +, --, --), where 0 indicates the drift direction of the first band is zero or unclear (Champion et al 2005), while the central two bands appear to reflect one another and the final band has a direction near-parallel to the third. In elliptical geometry the antisymmetry of the central bands suggests a sightline which crosses the beam close to the meridional line, while the lack of drift suggests a tilt in the sense shown in the top panel of Fig.~\ref{fig3} so that the line of sight is almost perpendicular to the elliptical trajectory and $\beta>0$. The fourth drift band would then (roughly) share the drift direction of the third.

McLaughlin {\etal} (2004) report that the repetition rate of each band is 15 rotation periods and in our simulation each beam was deemed to consist of $N=10$ emission regions, so the circulation time was $N\times{P_3}= 150P$ and, as can be seen in Fig.~\ref{fig3}, the structure of the circulation was set such that emission regions kept equal separation from one another. Together, these allowed the observed phase differences between the components to be reproduced (see middle panels of Fig.~\ref{fig3}). The remaining geometric parameters were largely constrained by the known width of the pulse window. In this simulation $\alpha=18\deg$ and $\beta=+0.2\deg$, the ellipses have eccentricity $e$ of 0.85, tilted at an angle $\Theta=+12\deg$ with outer and inner beam dimensions ($\rho_A, \rho_B$) of ($15\deg, 8\deg$) and ($6.5\deg, 3.4\deg$) respectively. Given the constraints imposed on the simulation, we consider the reproduction of the published single-pulse data, shown to the right of the simulation, to be satisfactory, displaying the appropriate drift directions (0, +, --, --) and observed phase differences between the components.

The integrated profile (shown in the bottom panel of Fig.~\ref{fig3}) is formed from a larger pulse sequence than shown in the middle panel and, although spaced in the pulse window as observed, the relative peak intensities differ somewhat from the published 325 MHz profile (McLaughlin {\etal}, 2004). It would be possible to further adjust the simulation so that, for example, emission was weaker in regions closer to the magnetic pole. However, we consider the discrepancy to be less important than a successful reproduction of the pulsar's observed subpulse pattern.    

\subsection{PSR B1839--04 (Q-mode)}

\begin{figure*}
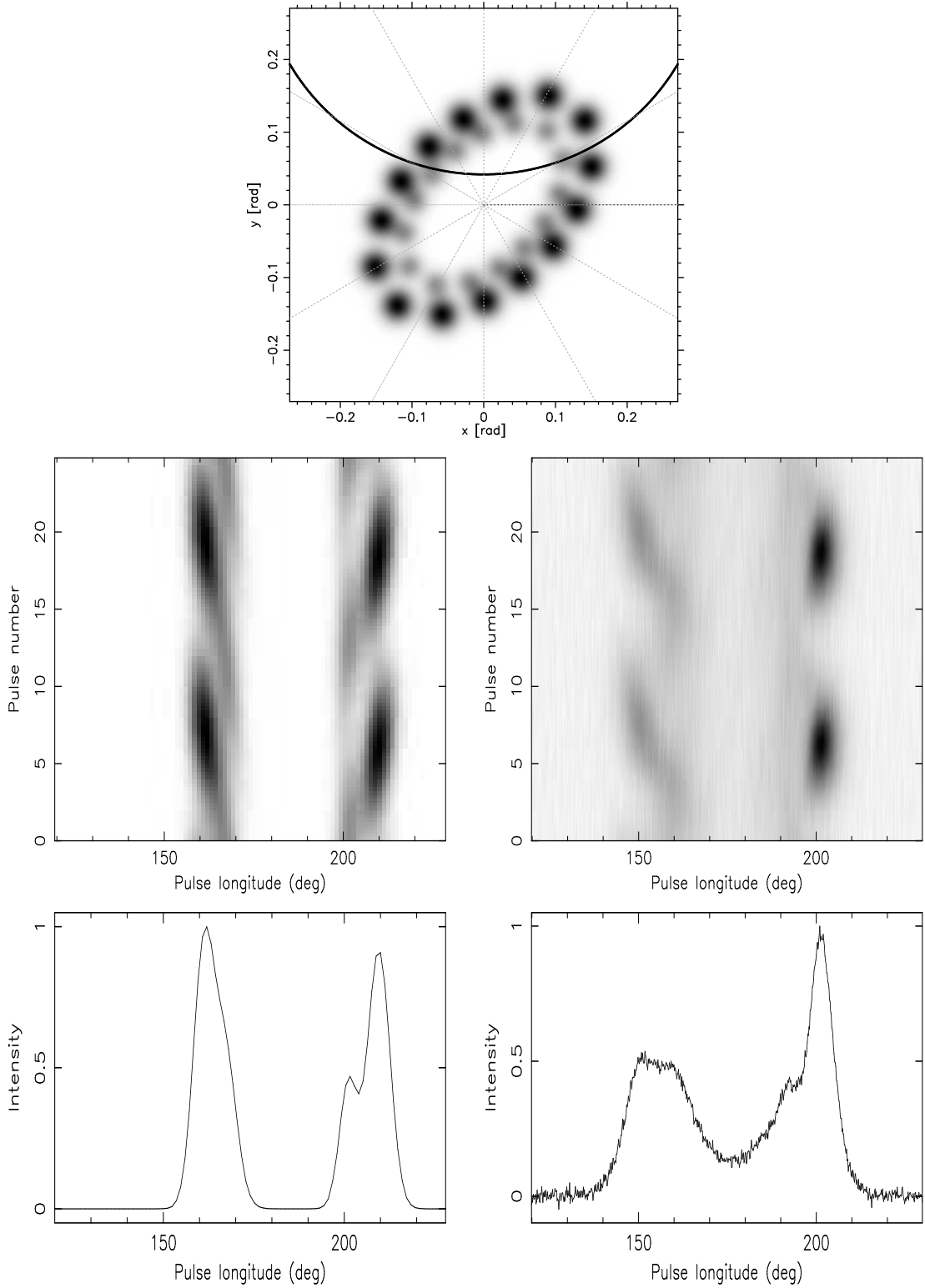

\begin{center}
\begin{tabular}{@{}l}
{\mbox{\includegraphics[width=7cm,height=7cm,angle=-90]{1839fakemap.ps}}}\\
\end{tabular}
\begin{tabular}{@{}l@{}l}
{\mbox{\includegraphics[width=7cm,height=7cm,angle=-90]{1839fakestack.ps}}}\ \ \ \ \ \ \ &
{\mbox{\includegraphics[width=7cm,height=7cm,angle=-90]{1839realstack.ps}}}\\
{\mbox{\includegraphics[width=6cm,height=7cm,angle=-90]{1839fakeprofile.ps}}}\ \ \ \ \ \ \ &
{\mbox{\includegraphics[width=6cm,height=7cm,angle=-90]{1839realprofile.ps}}}\\
\end{tabular}
\caption{A comparison of the simulation of the Q-mode emission of PSR B1839--04 with 1380 MHz Westerbork Synthesis Radio Telescope data as presented in paper I. \emph{top}: two nested elliptical beams with a common eccentricity, each containing 15 sub-beams tilted at $45\deg$ rotating anticlockwise about the magnetic axis, which is inclined at $20\deg$ to the pulsar's rotation axis and viewed with an impact parameter $\beta$ of $-2.4\deg$  (resulting in the line of sight indicated by the solid line). \emph{centre}: Emission drifts in opposing senses on either side of the fiducial plane (simulation left, observation right, both folded at the repetition period $P_3$). Note the shift in perceived central pulse longitude. \emph{bottom}: The resulting integrated profile (simulation left, observation right).}
\label{fig4}
\end{center}
\end{figure*}

The observed drift pattern (Paper I Fig. 7) of this pulsar's Q-mode emission is approximately  (--, --, 0, +), where the inner and outer beam components are effectively merged and denoted successively as Ia Ib IIa IIb. The drift rate is larger in the leading component than the opposing drift in the trailing component. The drift of the central components is less clear but drift is more discernible in the second component than the third (see Paper I Fig. 5) and has the same sense as the first. All this suggests an ellipse orientation as in the upper panel of Fig.~\ref{fig4}, where we have assumed $\Theta=-45\deg$ and $e=0.8$. The values taken for $\alpha$ and $\beta$ are $20\deg$ and $-2.4\deg$ respectively, consistent with allowed values derived from the PA swing (Paper I Fig. 8). 

To reproduce the observed single pulse sequence (Paper I Fig. 7) we adopted the observed $P_3=12.4P$ and assumed $N=15$ in each beam. The outer and inner beam sizes ($\rho_A, \rho_B$) were ($10.5\deg, 6.3\deg$) and ($7.7\deg, 4.6\deg$) respectively. As with PSR J0815+09, the emission regions were held at equal separations and the phase difference between the beams was set such that the emission regions of the inner beam were located at angles halfway between those of the outer beam. 

Despite artificial constraints, the middle panel of  Fig.~\ref{fig4} gives a good reproduction of B1839--04's Q-mode emission ({\cf} Paper I Fig. 7), delivering the observed drift patterns and subpulse phase shifts between the components. Note particularly the $10\degr$ shift between the observed centroid of the profile and location of the fiducial longitude implied by the simulation (at $180\degr$ per definition), so that the fiducial point lies at longitude $\sim170\degr$ of the observed profile. 

As with PSR J0815+09, the integrated Q-mode profile of the simulation has emission peaks with relative heights differing from those observed, and again we would argue that this could be the result of an intensity distribution across the beam (see Section \ref{SectPartialCone}) and does not suggest any underlying conflict between model and observation.

\section{The fiducial plane in a tilted geometry}
The fiducial plane is defined as the plane which contains the rotation and magnetic axes. When we observe a beam symmetric about this axis the pulse profile should have its mid-point exactly at the centre of the profile. In the Rotating Vector Model (Radhakrishnan \& Cooke 1969) the longitude of the maximum gradient in the polarisation position angle (PA) swing occurs at the fiducial plane, and any small (frequency dependent) deviation from this is usually interpreted as an aberration/retardation (A/R) effect (Blaskiewicz \etal, 1991).

\begin{figure*}
\begin{center}
\includegraphics[width=0.4\hsize,angle=0,trim=10 40 10 40,clip]{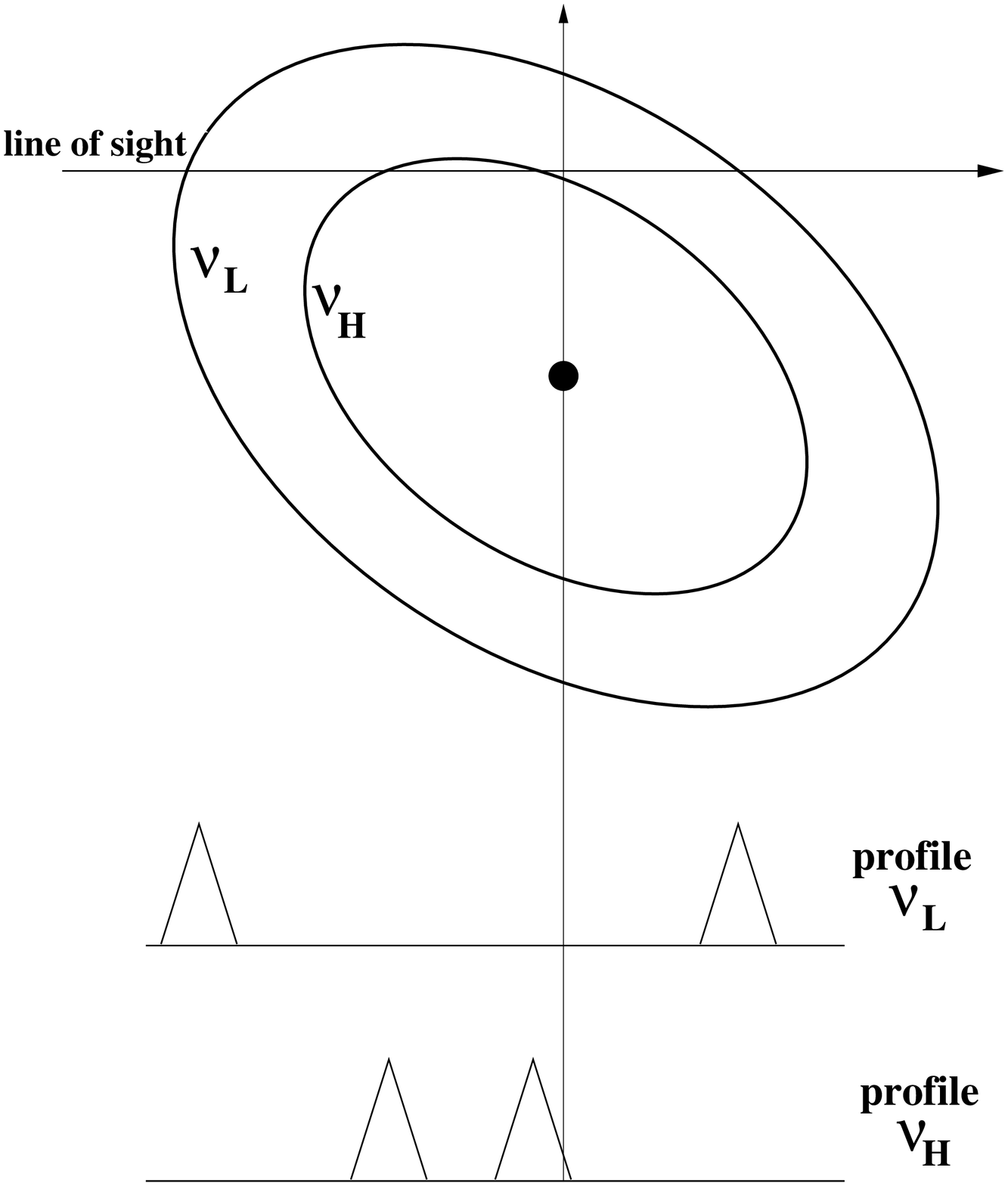}
\hspace{0.05\hsize}
\includegraphics[width=0.4\hsize,angle=0,trim=10 40 10 40,clip]{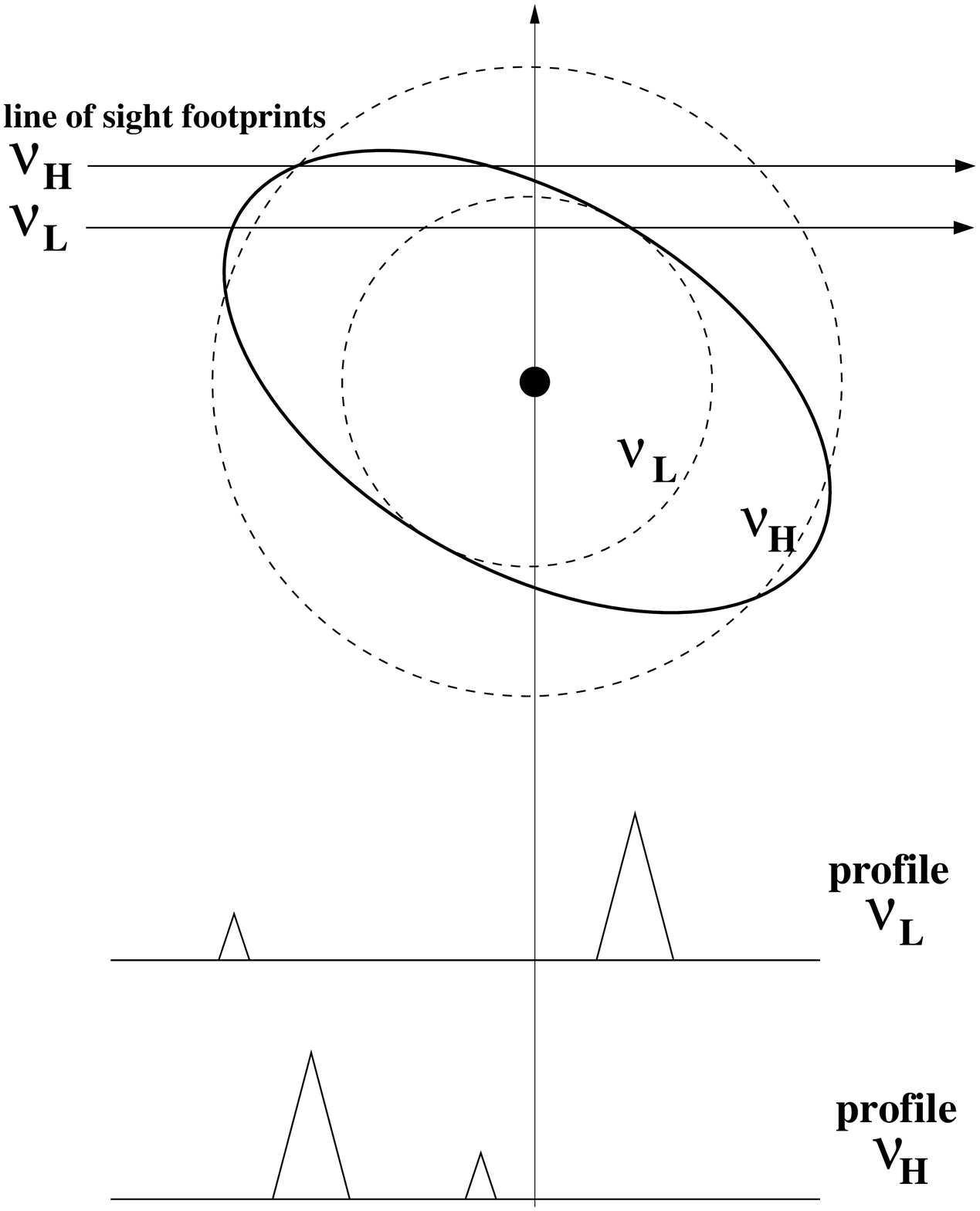}
\caption{An elliptical carousel tilted with respect to the (vertical) fiducial plane and centred on the magnetic pole is observed at a low ($\nu_L$) and high ($\nu_H$) frequency. Assuming a radius-to-frequency mapping the $\nu_L$ beam is produced at a higher altitude than the $\nu_H$ beam, and is wider as a consequence. Aberration/retardation effects are neglected. {\em Left:} A beam map, showing the observer's fixed line of sight intersecting the different sized beams at the two frequencies. This results in profiles which become not only narrow at higher frequencies but increasingly asymmetric about the fiducial plane, possibly in the most extreme cases with both components being observed ahead of the fiducial plane, as shown. {\em Right:} A polar cap map showing the same scenario projected onto the polar cap by considering the footprints of the fieldlines responsible for the observed beam, with a single fixed elliptical carousel forming beams at the two frequencies at different altitudes. The line of sight corresponding to frequencies $\nu_H$ and $\nu_L$ have different projections and are shown to generate the wide and narrow profiles respectively, as before. If the emission also depends on frequency such that fieldlines closer to the magnetic axis radiate more efficiently at low frequencies (the radial distance is indicated in the figure as dotted circles), then the profile components are not only asymmetric in phase longitude but also in intensity and have different spectral evolution, an effect which would not be seen in circular carousels.}
\label{fig2a}
\end{center}
\end{figure*}

In a tilted geometry we must be careful with our interpretation of both the intensity profile and its relation to the PA swing. The centroid of the profile will either lead or trail the fiducial plane, depending on the sign of the angle of tilt and the sign of $\beta$ (see Fig.~\ref{Fig1}, right, and Fig.~\ref{fig2a}, left). However, if $\beta=0$ the profile will be symmetric even if the carousel is extremely eccentric ($e\approx1$) and highly tilted ($\Theta\approx45\deg$). For higher $|\beta|$ the effect of tilt will be increasingly detectable, and the profile centroid will not be observed in the fiducial plane. In the case where $|\beta|$ is such that  the line of sight is grazing the beam (hence near-tangential to the edge of the beam) then the centroid is shifted to a phase longitude where the PA is offset by the angle $\Delta\psi_2$ (see Fig.~\ref{Fig1}, right) given by 
\eq
\lab{Eq5}
\tan\Delta\psi_2=\frac{e^2\sin\Theta\cos\Theta}{1-e^2\cos^2\Theta}
\en
(see Appendix). The observational result will be a single-humped profile displaced by a longitudinal phase of $\Delta\phi$ from the fiducial plane (the direction depending on the sign of $\Theta$). Assuming suitable values for $\alpha$ and $\beta$ derived from the shape of the PA swing, $\Delta\phi$ can be related to $\Delta\psi_2$ using the standard formula (Lorimer \& Kramer 2005)
\eq
\lab{Eq6}
\tan\Delta\psi_2=\frac{\sin\alpha\sin\Delta\phi}{\sin(\alpha+\beta)\cos\alpha-\cos(\alpha+\beta)\sin\alpha\cos\Delta\phi)}.
\en
Therefore a measurement of $\Delta\phi$, the offset between the pulse longitude corresponding to the maximum gradient of the PA swing and the midpoint of the profile, allows the $\Delta\psi_2$ to be determined. Using Eq. \ref{Eq5} this results in a constraint on the eccentricity and tilt of the elliptical beam. In addition, this effect offers a potential explanation for some pulsars with apparently ``missing'' components (Lyne \& Manchester 1988, Mitra \& Rankin 2011) and is discussed further in Section \ref{SectPartialCone}.

The effect of A/R on a profile at normal pulsar emission altitudes is to shift the profile to earlier longitudes and the PA curve by an identical amount to later longitudes ({\eg} see Fig. 2 of Dyks 2008). With a carousel tilted in the sense of Fig.~\ref{Fig1} a line of sight traversing above the $x$-axis ($\beta<0$), the centroid shift will $\emph{add}$ to the A/R effect, and below the $x$-axis ($\beta>0$) will counter it. A carousel tilted in the opposite sense to Fig.~\ref{Fig1} will exhibit opposite effects.

It was noted in Paper I (using data from Gould \& Lyne 1998) that the PA swing maximum in \mbox{PSR B1839--04} possibly $\emph{precedes}$ the fiducial point defined by the profile midpoint (although not significantly). This possibility would be supported by the tilt and line of sight shown in Fig.~\ref{fig4} (top), with the resulting shift in the fiducial point ($\sim10\degr$) sufficient to overcome and reverse any A/R effects. However, the observed profile midpoint of this pulsar is likely to be dominated by the weaker but more sustained Q-mode (see Fig. 2 of Paper I), which itself appears shifted earlier relative to the B-mode. So the profile midpoint of the bi-drifting Q-mode, if emitted at the same altitude as the B-mode, may be less shifted with respect to the PA swing maximum than in the total profile. Single-pulse observations of this pulsar with full polarisation would quantify and clarify this point. 

\subsection{\label{SectPartialCone}``Partial cone'' pulsars}
In general, the offset of the profile relative to the PA swing arising from a tilted beam can be expected to be relatively small in bi-drifting pulsars since these require $|\beta|/\rho$ to be small. However, since in the pulsar population at large the distribution of  $\beta$ is random, there may well exist pulsars observed at large $|\beta|/\rho$ which possess tilted beams although not exhibiting bi-drifting. In such pulsars the PA swing will be typically shallower and the observed PA swing offset greater than in the low $|\beta|/\rho$ case. We can therefore expect to find at least some pulsars with their steepest gradient of the PA swing strongly offset to earlier or later longitudes with respect to the profile as a result of their tilt. In extreme cases (large tilt, high eccentricity and large $|\beta|/\rho$) it is possible for the PA swing maximum (i.e. fiducial plane) to lie completely outside the phase longitudes of the profile components and may lead their beams to be categorised as ``partial cones''  (Lyne \& Manchester 1988). 

A tilted beam combined with radius-to-frequency mapping can bring about major frequency-dependent variations in the observed relative longitudes of the components in the pulsar's profile (see Fig.~\ref{fig2a}, left). At lower frequencies the line of sight will traverse a wide beam generated higher in the magnetosphere by fieldlines which lie closer to the pole on the polar cap. Its intersection with the elliptical beam will therefore generate components more symmetric about the fiducial plane, whatever the tilt of the ellipse. At higher frequencies the asymmetry will grow, with one of the components ``moving'' towards the fiducial plane faster than the other, possibly resulting in both components lying on the same side of the fiducial plane (Fig.~\ref{fig2a}, left). Simultaneous observations at multiple frequencies could then delineate a map of the elliptical beams. This effect might lie behind the apparently strange multi-frequency behaviour of the components of PSRs B1133+16, B1919+21, B0809+74 and others (see Hassall {\etal} 2012, 2013).

Fig.~\ref{fig2a}, right, presents the same information as in Fig.~\ref{fig2a}, left, but now projected onto the polar cap where the footprints of the fieldlines responsible for the beam form the underlying carousel and the low frequency ($\nu_L$) sightline projects more closely to the magnetic pole than that of the high frequency ($\nu_H$). If we additionally postulate that there is a frequency dependence on the footprint parameter $s$ (the radial distance indicated in the figure as dotted circles), such that for example fieldlines with small $s$ radiate more efficiently at low frequencies, then it can be seen that the separate components are not only asymmetric in longitude but also in intensity. In such a scenario it is not impossible that one or other of the components becomes near-invisible in the profile at a given frequency, a likelihood which will increase with tilt and eccentricity. This effect might explain the apparent absence of components either preceding or following the core emission noted by Lyne \& Manchester (1988). The work of Mitra \& Rankin (2011) showing that many of the ``missing'' components were in fact present, but simply much weaker than the other at a particular observing frequency, supports this view.

\begin{figure*}
\begin{center}
\begin{tabular}{@{}l@{}l}
{\mbox{\includegraphics[width=6.5cm,angle=-90]{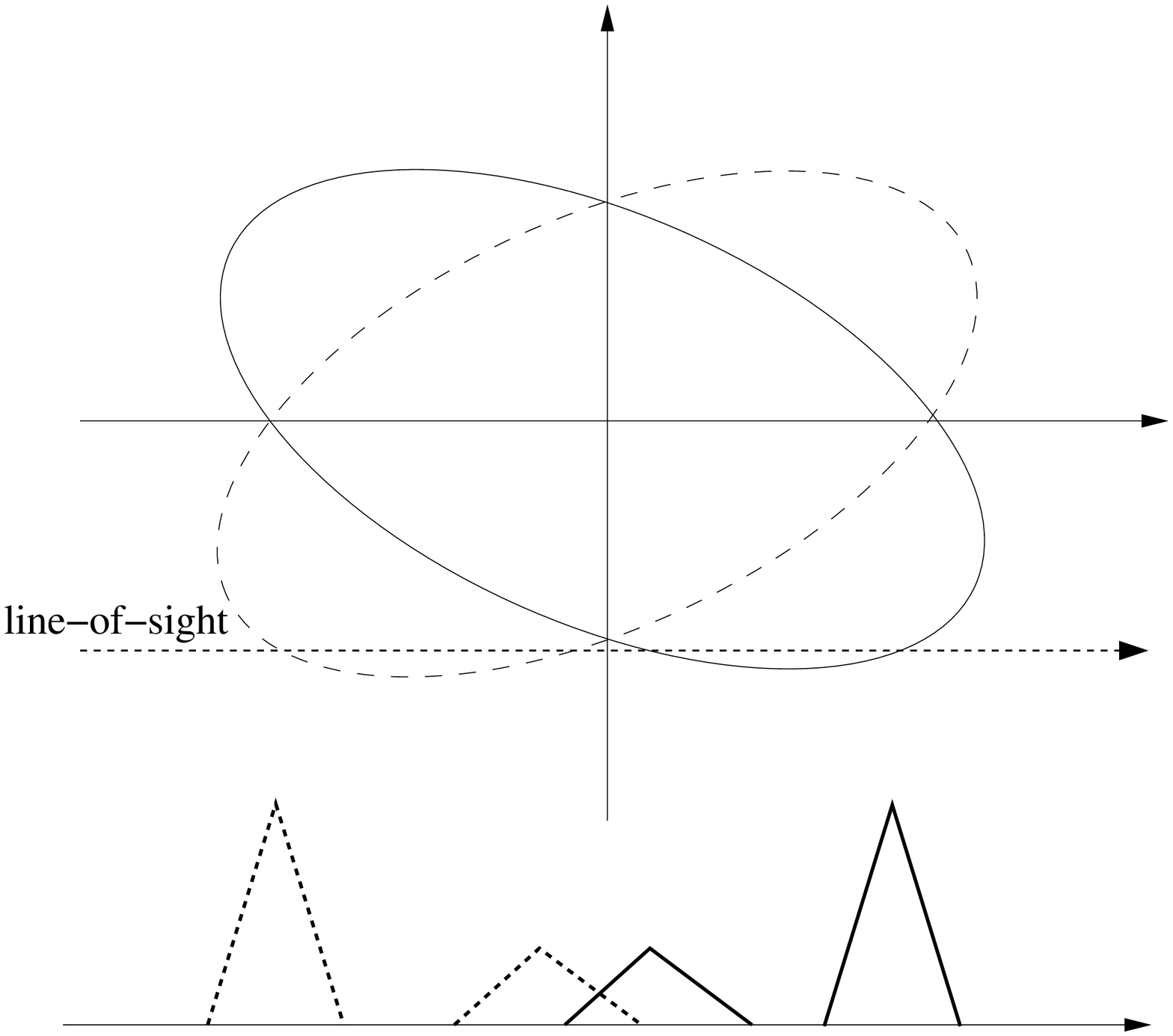}}}\ \ \ \ \ \ \ &
{\mbox{\includegraphics[width=6.5cm,angle=-90]{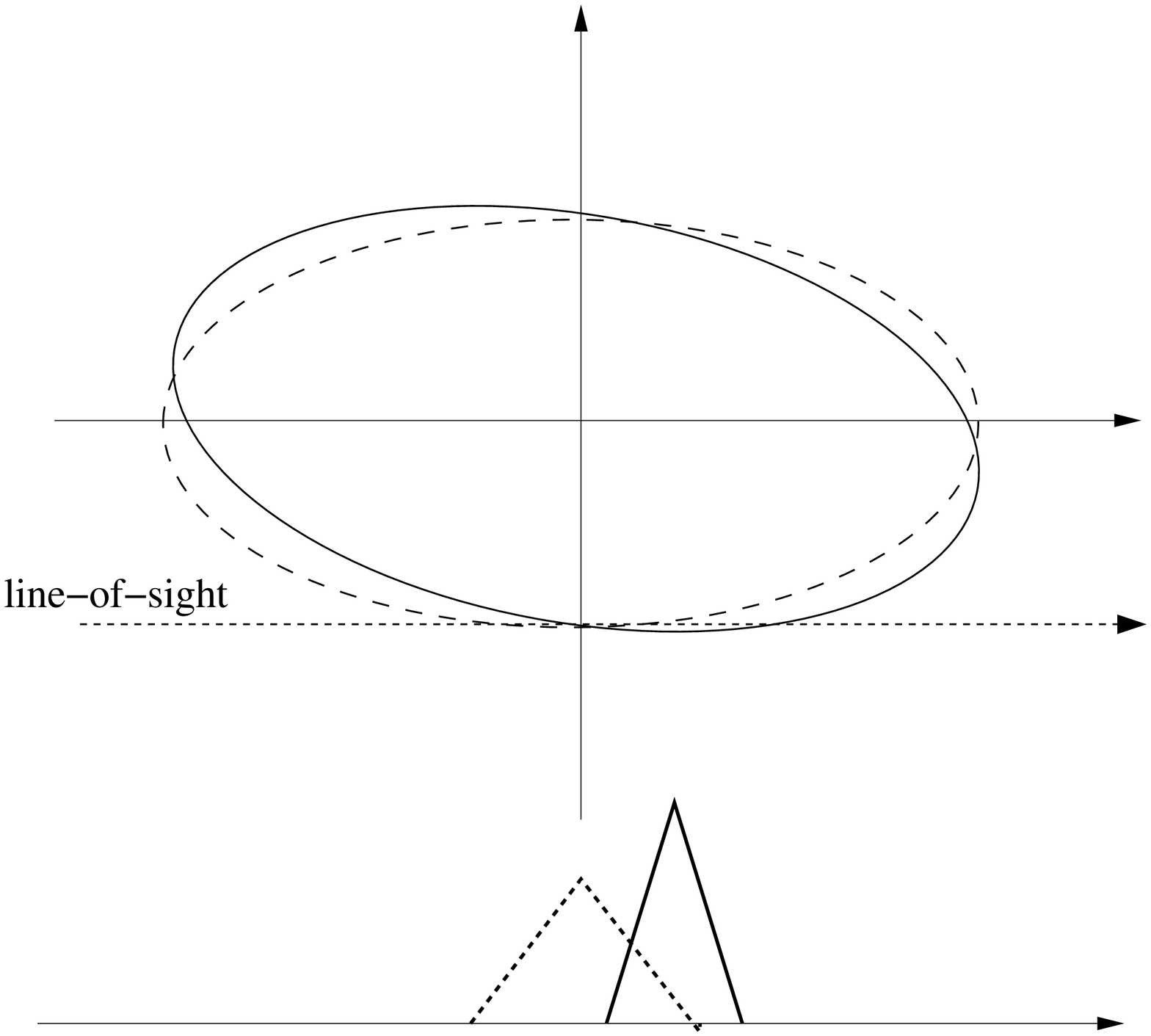}}}\\
\end{tabular}
\caption{Possible line of sight traverses for ``flare'' pulsars PSR B1859+07 (left) and PSR B0919+06 (right) and the resulting profile structures, showing the ``normal'' beams in dark lines and flare beams in dashed lines. The central meridian represents the fiducial plane.}
\label{Fig5}
\end{center}
\end{figure*}

Among the pulsars thought to have ``missing'' components are the so-called ``flare'' pulsars. On closer inspection some of these have been found to have weak leading/trailing features which may be generated by occasional bunched single pulse excursions to earlier/later longitudes (Mitra \& Rankin 2011). In the cases of PSRs B0919+06 and B1859+07 (Rankin {\etal} 2006, Wahl {\etal} 2016) the excursions appear to have a quasi-periodic pattern (Perera {\etal} 2015, 2016, Han {\etal} 2016). Some authors speculate that the ``missing'' components of these pulsars are the result of changes in absorption and/or altitude, but in the present context we suggest that the same effect may be produced by a fairly rapid (but not necessarily instant) change in tilt in their respective underlying carousels which would then be reversed a few pulses later (see Fig. \ref{Fig5}). This may be accompanied by a change in the frequency-dependence of the profile when the carousel crosses different fieldlines (see Fig.~\ref{fig2a}, left). A careful modelling of these pulsars including a fit of the observed PA swing and the suggested fiducial plane would be a test of the viability of this idea. 

It should be noted that the identification of tilted beams need not rely on polarisation properties. In PSR B2303+30 (Redman {\etal} 2006, Fig. 15) the subpulse separation ($P_2$) of the pulsar's drift  ($P_3\approx{2}P$) is observed to become narrower on the trailing edge of the asymmetric profile. Thus both profile and subpulse behaviour are asymmetric. The authors suggest that an underlying symmetric profile (with symmetry in $P_2$ also) may be present, but with its trailing part ``missing'' as a result of  magnetospheric ``absorption''. However, in the context of the present model, we may interpret the asymmetry as a consequence of the line of sight at the observing frequency (430 MHz) grazing the top or bottom of a tilted ellipse. Radius-to-frequency mapping suggests that an unequal double-peaked profile may become apparent at lower frequencies, where the sightline intersects and reveals more of the beam. 

\section{Alignment-varying beams}
The example of flare pulsars in the previous Section suggests exploring the possibility that emission beams may sometimes change their alignments on an observable timescale. A number of pulsars have gradually shifting profiles which may be interpreted in this way. These shifts usually occur after or before a null or mode-change and are often accompanied by a change in the profile width. Again, it is sometimes possible to interpret this as a change in altitude in the emission region, but we point out here that a small rotation in the orientation of a non-circular beams would have the same effect. Examples are the extreme nulling (93\% of the time) pulsar J1502--5653 (Li {\etal} 2012, see their Figs. 5 \& 7, which show a post-null adjustment and a widening profile),  and a similar but rare effect in the weakly nulling (1.4\%) pulsar B0809+74 in which nulls are occasionally accompanied by subpulse mode-changing (van Leeuwen {\etal} 2003). 

The case of PSR B0818--41 is particularly interesting (Bhattacharyya {\etal} 2010) because the pulses following a null not only shift to later longitudes, they are also \emph{wider}, which -- as those authors point out -- is counter to expectation if an emission altitude change alone were responsible. A further feature of this pulsar is that its drift patterns alter before and after nulls, something which might be expected if the carousel is changing orientation on a timescale comparable to that of the subpulse drift, so that with each pulse our line of sight would sample a different section of the carousel. 

A similar case is PSR B0031--07. This pulsar is known to possess three modes, each with a distinctive repetition rate $P_3$ (Huguenin {\etal} 1970). The profile widens as the drift rate increases (Wright \& Fowler 1981a), but each mode profile is centred at slightly later longitude (Vivekanand \& Joshi 1997). Again, this is not easily explained by changing emission altitude, so a slight but rapid change in orientation of the carousel is a possible explanation. Furthermore, within each mode the separation ($P_2$) of the drift bands can be seen to gradually widen (see Fig. 2 of Vivekanand \& Joshi 1997), which might be explained by a further slow adjustment of the orientation.

A further example is PSR B2319+60, which has two clear drift modes both producing a profile of which the trailing component (at 1415 MHz) is far stronger than the leading component.  These modes are frequently interrupted by an ``abnormal'' mode with irregular subpulse modulation, where now the leading half of the profile dominates. This profile behaviour is very similar to flare pulsar B1859+07 modelled in Fig.~\ref{Fig5} ({\cf} Fig. 2 of Wright \& Fowler 1981b), suggesting a common explanation.  

When a radius-to-frequency mapping (see Fig.~\ref{fig2a}) is combined with a changing carousel orientation the resulting mode switch will change the spectral evolution of the profile components. The recently-investigated emission of the interesting single-peaked pulsar B0611+22 (Seymour {\etal} 2014, Rajwade {\etal} 2016) shows that the sudden mode changes are both frequency and pulse phase dependent.  The pulsar has an asymmetric PA swing which apparently steepens on its trailing edge (Lyne \& Manchester 1988, Johnston {\etal} 2008). It may be possible to devise a model which sees the switch as resulting from a flip in an elliptical carousel whose fiducial plane is located on the trailing side of the profile. 

A general observational test of all these interpretations would be to see if the profile shift is accompanied by a shift in the longitude of the maximum gradient of the PA swing. If the profile shift is due to changing altitude, the PA swing will also shift due to the A/R effect. If the profile shift is due to a reorientation of the carousel at the same altitude, then the PA swing will not shift.

\subsection{PSR B0826--34}
The emission of the complex pulsar B0826--34 (Biggs {\etal} 1985, Gupta {\etal} 2004, Esamdin {\etal} 2005, Bhattacharrya {\etal} 2008, van Leeuwen \& Timokhin 2012) extends throughout most of its pulse. It is possible to discern up to 13 distinct driftbands across the profile and the behaviour of this pulsar is undoubtedly strong evidence for the existence of polar cap carousels. Its magnetic axis is almost certainly close to alignment with the rotation axis, so our line of sight gives us a head-on view of the polar cap. The widely-accepted view is that the pulsar possesses two nested circular carousels grazed by our circular line of sight traverse, and this gives rise to separate pulse regions.

However, there is an underlying problem with this model. Esamdin {\etal} calculate the dimensions of the carousels based on the varying $P_2$ separations of the subpulses. This suggests an $\alpha$ of just $0.5\degr$ and $\beta$ of $\sim4.5\degr$, giving $\alpha/\beta\approx 0.11$. Gupta {\etal} (2004) base their model on the much larger ratio $\alpha/\beta\approx 2.2$,  which they derive from the PA swing maximum gradient observed by a number of authors (Biggs {\etal} 1985, Lyne \& Manchester 1988, Rankin 1993). This leads them to conclude that the model can only be made compatible with the nested carousel picture if the common centre of the carousels is located some degrees away from the magnetic pole, a feature not found in other double-cone examples (PSRs B1237+25 and B0818--41). Furthermore, the pulsar's polarisation PA swing is strongly displaced from from the centroid of the main pulse profile and this is not easy to account for. 

We suggest that these modelling difficulties stem from the assumption that the carousels are circular. While fitting the complex data of this pulsar is undoubtedly a difficult task, it may be worth looking at an alternative picture of an elliptical carousel (possibly just one, rather than two), and perhaps not precisely aligned with the fiducial plane (to explain the shift in PA swing). In the pulsar's weak mode (Esamdin {\etal} 2012) the carousel's orientation may change. Dropping the requirement for circularity would give greater freedom to the modelling effort and potentially make the pulsar's behaviour compatible with many of those described elsewhere in this paper.

\subsection{The B and Q modes of PSR B0943+10}
PSR B0943+10 has been the subject of intense study over the last 20 years ({\eg} Deshpande \& Rankin 2001, Suleymanova \& Rankin 2009, Backus {\etal} 2011, Bilous {\etal} 2014). Like PSR B1839--04, it has a mode exhibiting highly regular drift which switches rapidly to and from a chaotic mode, and both pulsars are thought to have low inclinations between their magnetic and spin axes. The proportion of time spent in their chaotic modes is roughly comparable in the two pulsars ($\sim30\%$ of the time in the B-mode of PSR B0943+10  and $\sim 25\%$ in the Q-mode of PSR B1839--04) and their spin periods $P$ are not dissimilar (1.1 s versus 1.9 s). However, their subpulse modulation timescales are very different: their $P_3$ values are respectively $1.86P$ and $12.4P$, and while in PSR B0943+10 the modes typically last several hours, in PSR B1839--04 the mode durations are of the order of a few minutes.

In PSR B0943+10 it is known that the B to Q change is accompanied by the appearance of a precursor component, a dramatic increase in thermal X-ray emission (Hermsen {\etal} 2013, Mereghetti {\etal} 2016) and chaotic subpulse behaviour. It has recently been found (Bilous {\etal} 2014) that as the B-mode progresses the centroid of that mode's double-peak profile shifts markedly to later longitudes and the strength of the trailing peak diminishes. Immediately following the change to the Q-mode the subpulse behaviour becomes chaotic and the profile reverts to a single-peak profile whose centroid is at the original longitude of the B-mode centroid. 

Interpreting this result in geometric terms, it is possible that the B-mode beam of PSR B0943+10 is elliptical in shape and initially symmetric about its fiducial plane. In the course of the B-mode, the elliptical beam with its highly ordered sub-beams gradually tilts until it reaches a limiting inclination. This triggers a rapid transformation to a beam which is chaotic but again symmetric about the fiducial plane. In support of this tilted model ({\eg} Fig.~\ref{fig2a}) is the observed gradual change in the ratio of the intensities of the two components, and the fact that this ratio is different at different frequencies (Backus {\etal} 2011, Bilous {\etal} 2014). Note, however, that this picture is the reverse of what we suppose is happening in PSR B1839--04, where the chaotic mode would appear to tilt the ellipse even further, so without carefully identifying the fiducial plane we should not prescribe the tilt direction of PSR B0943+10.

During the B-mode sequences of PSR B0943+10 the profile shift is accompanied by a gradual decrease in $P_2$ (Backus {\etal} 2011), which is plausibly argued to result from an equivalent exponential increase in the circulation time of the carousel's sub-beams from $36.5P$ to $38P$ and an increase in their average number from 18 to 20. However, behind this argument is the assumption that the carousel is circular.  It is possible that the observed reduction in $P_2$ may be the consequence of the gradual change in our sightline's intersection of the beam as the carousel tilts, so that no change in the number of sub-beams would be necessary -- although the tilting in itself would not necessarily generate change in the circulation time.  Clearly, this line of argument remains vague until a detailed model is attempted, but could have interesting implications for interpreting the apparent evolution of pulsed X-ray emission during the B-mode (Mereghetti {\etal} 2016).
  
\section{Physical interpretation}
Explaining drifting subpulses by ${{\bm E}\times {\bm B}}$ drift on the pulsar's polar cap was the groundbreaking idea of Ruderman and Sutherland in 1975. In the the original model the polar cap and the drift were assumed perfectly circular as a reasonable first approximation. Difficulties arose when the observed drift rates did not match predictions and the model was modified with plausible additional polar cap physics (Gil \& Sendyk 2000, Gil {\etal} 2003) so that the surface potential differences were partly screened. 

The complex profiles of many pulsars were evidence that the emission beams took the form of single or double cones (sometimes plus core emission), and these were assumed to be circular (Rankin 1990, 1993). To explain their presence (and the separation of drifting subpulses) it was argued, again reasonably, that the polar cap potential distribution caused sparks to develop at  discrete distances from one another.

However, none of this requires that the spark trajectories across the polar cap should be precisely circular. An important physical feature of the polar cap model is that multipoles should exist close to the surface in order to facilitate primary pair production, so surely these must cause irregularities in the spark trajectories. Furthermore, it is not clear why the trajectories of the sparks should repeatedly execute the same closed paths. Indeed, some have felt that the emission must be more ``patchy'' (Lyne \& Manchester 1988).

Arbitrarily non-circular trajectories of sparks were already discussed by Jones 2013 and paper I in the context of bi-drifting. In this paper we explore the concept of non-circular closed carousels further to explain a range of observed phenomena. This is a small step away from the rigidity of circular paths, but raises many questions. Among these are the interpretation of the polarisation PA swing in an asymmetric beam and the identification of the fiducial plane. We have also suggested that changes in a pulsar's profile may indicate that the underlying trajectory on the polar cap has changed (rather than emission becoming ``absorbed'' or shifted by a change in emission height).  Some authors have attempted to explain mode changing through physical changes on the polar cap (Zhang {\etal} 1997, Jones 2013, Szary {\etal} 2015). However, it is still necessary to explain how carousels can have distinct repeating closed configurations, whether circular or not, defined solely by conditions on the polar cap.

Alternatively, we may abandon the idea that the observed single pulse sequences and profiles are exclusively driven by events on the polar cap, and investigate feedback from the magnetosphere (possibly in the form of the return current).  There is powerful evidence for magnetospheric connection from the observations of interpulses in PSRs B1822--09, B1702--19 and B1055--52 (Fowler \& Wright 1982, Gil {\etal} 1994, Weltevrede {\etal} 2007b, 2012), where the opposing polar caps are clearly interacting with each other. Wright (2003) has argued that the inner and outer cones may be formed by interactions between the polar cap and the extremes of the magnetosphere's outer gap. 

A number of authors have discussed mode-changing as a sudden switch in the magnetospheric state (Goodwin {\etal} 2004, Timokhin 2010, van Leeuwen \& Timokhin 2012, Kalapotharakos {\etal} 2012, Jones 2012). In these calculations the separate states are broadly connected to polar cap conditions, but still lack predictive power for specific pulsars. Recently, however, the mode change in the radio emission of PSR B0943+10 has been found to coincide with a change in X-ray emission from that pulsar, and this suggests that mode-changing is indeed a pan-magnetospheric event (Hermsen {\etal} 2013, Mereghetti {\etal} 2016).   

A more speculative approach might be to consider the presence of planetary/asteroidal bodies in orbits about the pulsar (Cordes \& Shannon 2008, Wahl {\etal} 2016). These might be expected to generate precise periodic changes in the pulsar emission, but so far most mode-changing and nulling seems at best ``quasi-periodic''. Interestingly, it can be shown that the polar cap footprints of circular orbits would take the form of ellipses with eccentricity $\sin\alpha$. However, the creation of a tilted ellipse would require the orbit not to lie in the neutron star's equatorial plane.

\section{Summary}
Bi-drifting is a rare phenomenon, only two clear examples being identified among the $\sim{100}$ known pulsars with regular subpulse drift ({\eg} Weltevrede et al 2006, 2007, Basu {\etal} 2016).  Nevertheless, it has always been felt that the lack of explanation for this phenomenon has placed a significant question mark over the classical model for drifting subpulses presented by Ruderman \& Sutherland (1975) and later developed by others ({\eg} Gil {\etal} 2003).

Here we propose a simple geometric solution by relaxing the condition that the observed subpulses are following a circular path as they drift across the pulse window\footnote{Note that Perera {\etal} 2010 find it necessary to introduce a highly elliptical beam, albeit symmetrically aligned, to explain the precession-driven profile of PSR J0737--3039B in the double pulsar system.}. By considering an elliptical beam with a degree of tilt with respect to the fiducial plane it is shown that bi-drifting can be accounted for, both in single and double emission cones. 

A necessary (but not sufficient) condition for bi-drifting is that the line of sight passes close to the magnetic pole (low $|\beta|/\rho$). Assuming radio-to-frequency mapping, it may be found at low frequencies, but not at higher, in the same pulsar. We demonstrate on geometrical grounds that bi-drifting is statistically unlikely to be observed unless the tilt and eccentricity are extreme. 

Despite imposing the artificial constraint of requiring both inner and outer cones to be ellipses with common eccentricity and orientation of their major axes, it was possible to obtain plausible simulations for the bi-drifting pulsars B1839--04 and J0815+09. Both are relatively close to alignment and therefore the line of sight samples a curved path on the polar cap. These simulations schematically reproduce the observed drift rates and the phase differences from component to component. In reality, the carousel/beams need not be a formal mathematically-defined ellipse, but of more general convex oval form and perhaps not even centred on the magnetic pole.

Bi-drifting will only be observed when our line of sight by chance follows a very exceptional path. However, other lines of sight across tilted asymmetric beams, while producing the same drift direction in all components of the pulsar's profile,  will result in observing different drift rates from component to component (the same $P_3$ but different $P_2$). We note that this effect is seen in many pulsars and usually manifests itself in differing component widths.

An important expectation of the tilted model is that the fiducial plane will be shifted relative to the profile mid-point (and this could be in either direction). Any pulsar with a tilted carousel will exhibit this property whether or not it also shows bi-drifting, and the effect will be greater for pulsars whose beams are observed with large $|\beta|/\rho$. In fact, the degree of shift with respect to the polarisation PA curve can be used to constrain the angle of tilt and eccentricity. If radius-to-frequency mapping is present, then in a tilted geometry the pulsar's separate components can be expected to have differing spectra and their respective separations from the fixed fiducial plane will also be frequency-dependent and asymmetric. Many pulsars which appear to have ``missing'' components may be explained by some combination of these geometric effects.

We take the model further by suggesting that a range of observed phenomena can be understood if we allow the carousel to vary its orientation on the polar cap. This re-orientation may take different manifestations:  (a) the switch to a new mode, which is then sustained for many pulses ({\eg} PSRs B1839--04, B0943+10 and B0611+22), (b) the ``flares'' of certain pulsars, where the switch is less rapid but the new state is only briefly maintained ({\eg} PSRs B1859+07 and B0919+06), (c) the gradual drift of a profile centroid ({\eg} the Q-mode of PSR B0943+10) and (d) apparent variations in $P_3$ close to a null ({\eg} PSR B0818--41).  

Our model of an asymmetric carousel/beam, which began as a device for explaining bi-drifting, has been found to have wider applications and significance. It gives support to the original idea that polar cap carousels generated by ${\bm{E}\times {\bm B}}$ drift can explain the common pulsar phenomenon of drifting subpulses. On the other hand it has raised questions as to the identity of the carousel: why does it have a closed convex form? If this form can change (in a mode-change, for example), why does it apparently flip between the same two forms? Is the carousel defined exclusively by conditions on the polar cap or is it at least partly imposed by feedback from the magnetosphere?  No easy answer to these questions here, but we hope to have suggested a novel way of looking at the problem which will stimulate further tests of the carousel model.

\section{Acknowledgements}
The authors are grateful to Maura McLaughlin for permission to select the extract from the Arecibo observations of PSR B0815+09 which appears in Figure \ref{fig3}. GW wishes to thank to Manchester University for granting Visitor status. 
This work was supported by the UK Science and Technology Facilities Research Council (STFC), under grant number ST/L000768/1. The Arecibo Observatory is operated by SRI International under a cooperative agreement with the National Science Foundation (AST-1100968), and in alliance with Ana G. M\'endez-Universidad Metropolitana, and the Universities Space Research Association.

{}
\appendix
\section{A tilted ellipse in Cartesian coordinates}
Fig.~\ref{Fig1} (left and right) shows elliptical beams whose major axes are both tilted by an angle $\Theta$ to the $x$-axis. In Cartesian coordinates the equation of each can be written as
\begin{eqnarray}
\lab{EqA1}
\nonumber\left(\frac{\cos^2\Theta}{\rho^2_A}+\frac{\sin^2\Theta}{\rho^2_B}\right)x^2+\left(\frac{\sin^2\Theta}{\rho^2_A}+\frac{\cos^2\Theta}{\rho^2_B}\right)y^2\\
-2\sin\Theta\cos\Theta\left(\frac{1}{\rho^2_B}-\frac{1}{\rho^2_A}\right)xy=1.
\end{eqnarray}
In Fig.~\ref{Fig1} (left) the dotted line  $y=-D$ and the angle $\Delta\psi_1$ define the point where a tangent to the ellipse is vertical.  To derive $\Delta\psi_1$ we differentiate the above equation with respect to $y$ and set $dx/dy=0$ to obtain
\eq
\lab{EqA2}
\tan\Delta\psi_1=\frac{y}{x}=\left(\frac{1}{\rho^2_B}-\frac{1}{\rho^2_A}\right)\frac{\sin\Theta\cos\Theta}{\frac{\sin^2\Theta}{\rho^2_A}+\frac{\cos^2\Theta}{\rho^2_B}}.
\en
Using $\rho^2_B=\rho^2_A(1-e^2)$ this can be rewritten as
\eq
\lab{EqA3}
\frac{y}{x}=\frac{e^2\sin\Theta\cos\Theta}{1-e^2\sin^2\Theta}.
\en
Substituting for $y$ in equation \ref{EqA1} from \ref{EqA3} we obtain the $x$-coordinate of the tangent point as
\eq
\lab{EqA4}
x=\rho_A(1-e^2\sin^2\Theta)^{\frac{1}{2}}
\en
so finally that
\eq
\lab{EqA5}
D=x\tan\Delta\psi_1=\rho_A\frac{e^2\sin\Theta\cos\Theta}{(1-e^2\sin^2\Theta)^{\frac{1}{2}}}.
\en
Fig.~\ref{Fig1} (right) shows the maximum possible impact parameter $\beta_0$ as a horizontal tangent to the beam, with its grazing point subtending a position angle $\Delta\psi_2$ to the fiducial plane $x=0$. To determine $\beta_0$ and $\Delta\psi_2$ we differentiate equation \ref{EqA1} with respect to $x$ and set $dy/dx=0$  to obtain
\eq
\lab{EqA6}
\tan\Delta\psi_2=\frac{x}{y}=\left(\frac{1}{\rho^2_B}-\frac{1}{\rho^2_A}\right)\frac{\sin\Theta\cos\Theta}{\frac{\cos^2\Theta}{\rho^2_A}+\frac{\sin^2\Theta}{\rho^2_B}}.
\en
Then, in the same manner as equation \ref{EqA4}, \ref{EqA6} reduces to
\eq
\lab{EqA7}
\frac{x}{y}=\frac{e^2\sin\Theta\cos\Theta}{1-e^2\cos^2\Theta}.
\en
Substituting for $x$ from equation \ref{EqA7}, the $y$-coordinate ($\beta_0$) of the tangent point is given by
\eq
\lab{EqA8}
\beta_0=\rho_A(1-e^2\cos^2\Theta)^{\frac{1}{2}}.
\en
Note that equation \ref{EqA3} is identical to \ref{EqA7} under the transformation $\Theta\rightarrow(90\degr-\Theta)$ giving this symmetry to $\Delta\psi_1$ and $\Delta\psi_2$. For given eccentricity $e$ their maxima occur at maximum tilt $\Theta=45\degr$ and are given by
\eq
\lab{EqA9}
\Delta\psi_{1,2}=\arctan\frac{e^2}{2-e^2}.
\en

\bsp

\label{lastpage}

\end{document}